\documentclass[a4paper,20pt]{article}
    \usepackage[T1]{fontenc}
    \usepackage{graphicx}
    \usepackage{epsfig}
    \usepackage{amsmath}
    \usepackage{amsfonts}
    \renewcommand{\abstract}{}
\begin{document}
\makeatletter
\renewcommand{\@oddhead}{\textit{Advances in Astronomy and Space Physics} \hfil \textit{B. Kapanadze}}
\renewcommand{\@evenfoot}{\hfil \thepage \hfil}
\renewcommand{\@oddfoot}{\hfil \thepage \hfil}
\fontsize{11}{11} \selectfont

\title{X-ray Selected BL Lacertae Objects: Catalogue and Statistical  Properties }
\author{\textsl{B.\,Z.~Kapanadze$^{1}$}}
\date{}
\maketitle
\begin{center} {\small $^{1}$ Abastumani Astrophysical Observatory at the Ilia State University, 0162, Cholokashvili Av. 3/5, Tbilisi, Georgia\\
bk@genao.org}
\end{center}

\begin{abstract}
This talk focuses on the statistical properties of  X-ray selected
BL Lacertae objects (XBLs) whose catalogue has been compiled. It
consists of 312 sources from different X-ray surveys, unambiguously
identified to  mid-2010. Results of  the statistical research of
different observational quantities (redshift, muliwavelength
luminosities, host/nucleus absolute Magnitudes, central black hole
masses, synchrotron peak frequencies, broadband spectral indices)
are also provided and existence of the correlation between them is
proved.  Overall flux variability shows an increasing trend towards
greater frequencies. XBL are found to be much less active in point
of intra-night optical variability compared to radio-selected BL
Lacs (RBLs).  A separate list of 106 XBL candidates is also created
including the same characteristics for each source as in the case of
XBL catalogue.
\end{abstract}

\section*{Introduction}
\indent \indent BL Lacertae objects form one of the extreme subclass
of active galactic nuclei (AGN). The prototype of these sources, BL
Lacertae, was detected by \cite{H29} who classified it as a
short-period variable star of 13-15  stellar magnitude and named the
object as "363.1929 Lac". The name  "BL Lacertae" was given by van
Schewick in 1941. On the basis of the photographic plates, taken at
the Sonnenberg observatory during 1927-1933, he deduced that there
was an irregular variable star that varied between 13.5 mag and 15.1
mag (see \cite{B00}). After almost three decades, \cite{S68}
reported that BL Lacertae coincided with the radio source VRO
42.22.01. This was followed by the detection of high and variable
radio/optical polarization (\cite{M68}, \cite{V69}), rapid optical
variability with 0.1 mag over a few hours  and flicker with the
amplitude $\Delta V=0.03$ mag per 2 minutes (\cite{R70}), steep
optical spectrum following a single power law, similar to quasars
but showing no emission lines (\cite{O69}).

 \indent \indent On the basis of the absorption features, detected in the optical spectra obtained with the 5 m
 Hale telescope, \cite{O74} determined the redshift of BL
 Lacertae. The obtained value ($z=0.07$) revealed that there was an
 extragalactic source, hosted by an elliptical galaxy. During
 1970-1978, one detected up to 30 sources with the same properties. On the "Pitsburgh Conference on BL Lacertae Objects"
 (1978), \cite{B78} suggested that the extreme
 properties of these objects should be caused by a Doppler-boosted
 emission pointed to the observer. During this conference, Ed
 Speigel introduced a term "blazars" to denote an independent class
 of the extragalactic sources, including BL Lacertae objects (BLLs) and flat
 spectrume radio-quasars (FSRQs, showing the same features with
 additional presence of emission lines). X-ray satellite \emph{Einstein}
 and \emph{Energetic Gamma-Ray Experimental Telescope}
 (EGRET)introduced new eras in the investigation of these sources.
 It was revealed that blazars constitute the most observed class of
 the extragalactic sources through $\gamma-$rays  up to the  TeV frequencies.\\
 \indent \indent BLLs are thus the extragalactic sources with following features:\\
 a. quasi-featureless spectra, leak of the prominent emission  lines;\\
 b. strong radio sources with a core-dominant morphology;\\
 c. strong  and variable optical/radio polarization;\\
 d. violent flux variability through all spectral bands;\\
 e. observed superluminal motions;\\
 f. broad continuum extending from the radio to very high-energy $\gamma$-rays.

 \indent \indent Up to now,  more than 900 BLLs are unambiguously identified (see the second edition of the Roma-BZCAT,
 http://www.asdc.asi.it/bzcat). Bulk of them are originally detected either through radio or X-ray
 bands. Due to this reason, One divided them broadly into the
 radio-selected (RBLs) and X-ray selected ones (XBLs). However, these
 subclass differ each from other by their spectral energy
 distributions. A BLL is assumed to be a XBL if (\cite{W94})
\begin{equation}\label{form1}
log F_\emph{x}/F_r \geq-5.5
\end{equation}
where the X-ray flux density is taken at 1keV (2.42$\times10^{17}$
Hz) and the radio one - at 5 GHz (both in Janskys). We deal
otherwise with a RBL source.

\section*{Catalogue}

The XBL catalogue consists of 312 sources (see \cite{K11}),
consisting  of their equatorial coordinates, redshifts,
multiwavelength flux values and isotropic luminosities,
X-ray-to-radio flux ratios etc.  It was compiled on the basis of following X-ray surveys:\\
 1. \textbf{HEAO-1 Large Area Sky Survey} (\cite{W84}, 16
 sources).\\
2. \textbf{Einstein Observatory Medium Sensitivity
Survey}(\cite{M82}), \cite{G84};  25 sources);\\
 3. \textbf{Enstein Observatory Extended Medium Sensitivity Survey} (\cite{G90},42sources);\\
4. \textbf{Exosat High Galactic Latitude Survey} (\cite{G91}, 10
 sources);\\
 5. \textbf{Einstein Slew Survey} (\cite{E92}, 61 sources);\\
6. \textbf{ROSAT All-Sky Survey} (\cite{B92}, 306 sources);\\
7. \textbf{XMM-Newton Bright Serendipitious Survey} (\cite{D04}, 3
sources).\\
\indent \indent A great deal of the XBLs belongs to the different
surveys.
 A separate list of 106 XBL candidates is also created including the same characteristics for each source as in the case of XBL catalogue. We can not consider them as BLL sources mainly due to
lack of the optical spectroscopy. Their spectra are either not
published or they are of bad quality and we thus can not exclude the
existence of the emission lines. Their observational features are,
in fact, the same as for confirmed BLLs and we expect that  high
signal-to-noise ratio spectroscopy will boost the number of XBLs.

\begin{figure*}[!ht]
\centering
  \includegraphics[width=20cm, height=0.45\textheight]{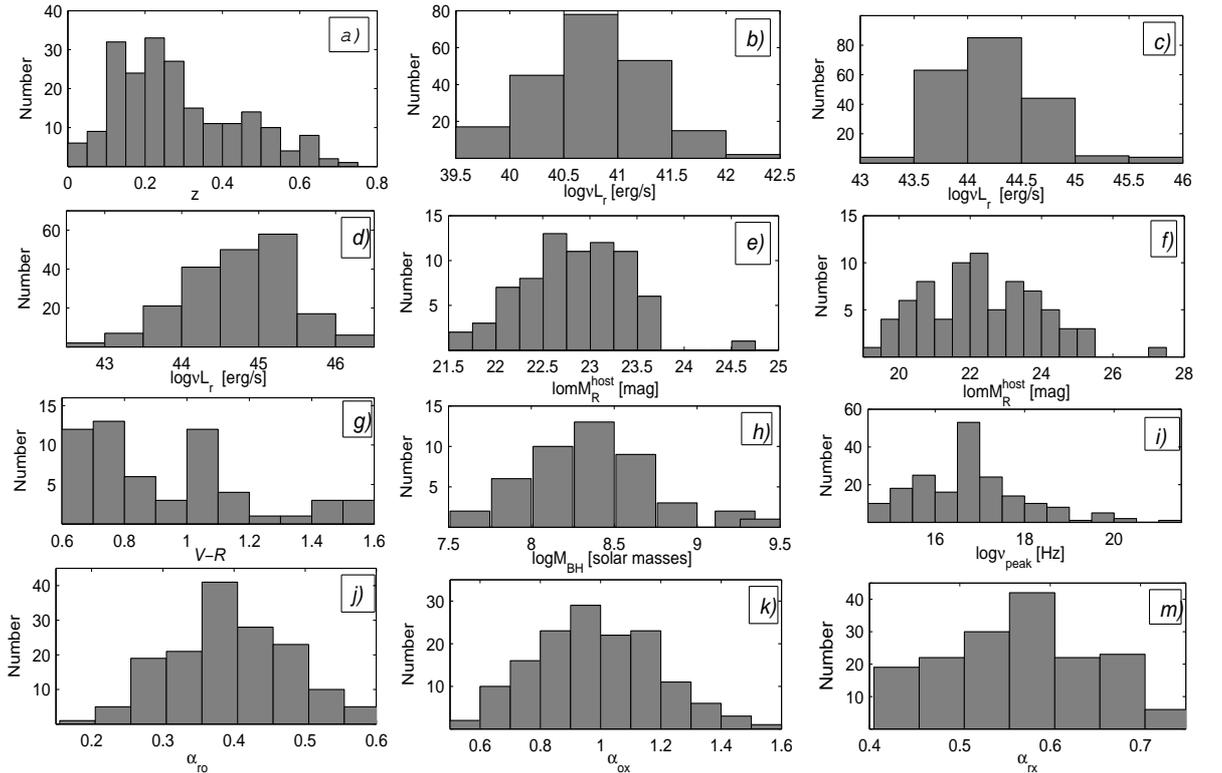}
 \caption{ \small Distribution of the observational quantities of XbLs.}\label{Fig1}
 \end{figure*}

\section*{Statistical Properties of XBLs}

Figure 1 gives the distributions of different observational
quantities: redshift, 1.4 GHz/V-band/ROSAT-band luminosities, host
and nucleus R band absolute magnitudes, host V-R indices etc.
Correlations between different quantities are given in Figure 2.

\begin{figure*}[t]
 \centering
 \includegraphics[width=20cm, height=0.45\textheight]{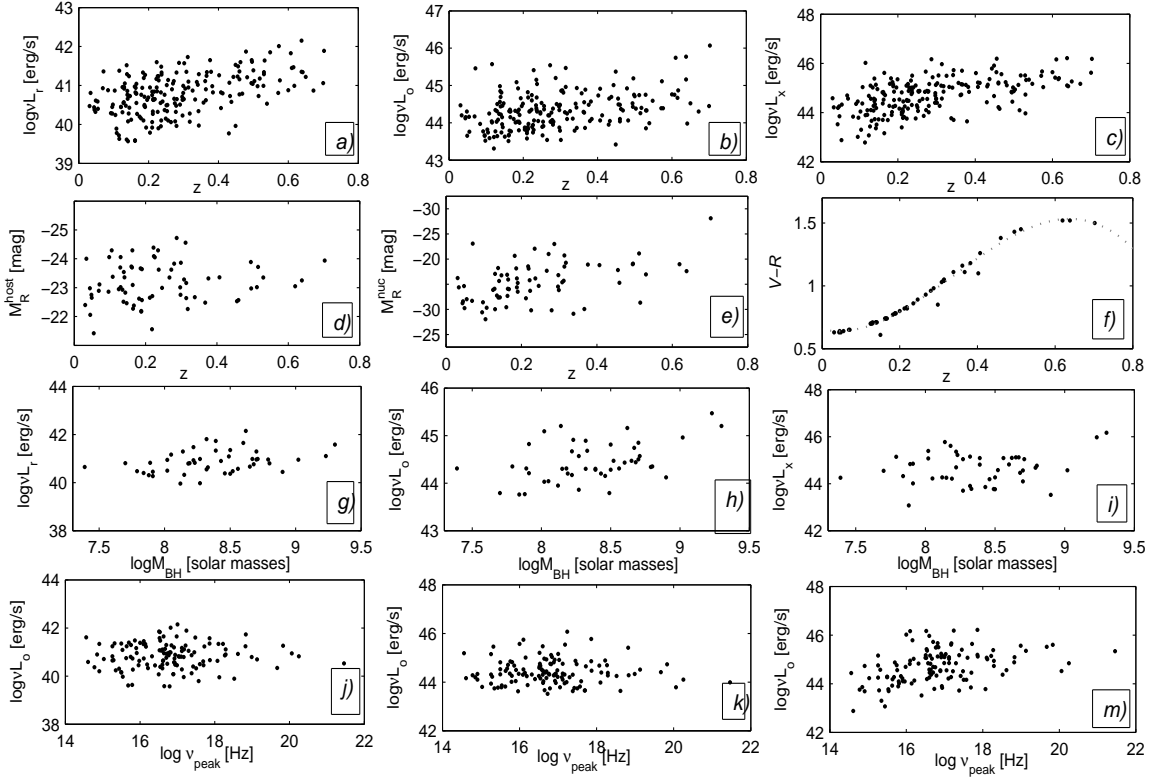}
 \caption{\small Correlation between the different observational quantities of XBLs.}\label{Fig2}
 \end{figure*}

\textbf{Redshifts}:only 207 XBLs (66.3\% of total sample) have
confirmed redshifts. They range from z=0.031 (Mrk 421) to z=0.702
(H151+660) with a peak of the distribution at z=0.23 (Fig1a). About
37\% of the sources are concentrated within z=0.1 - 0.3, and 88\% of
the sources have z<0.5. Redshifts of 105 sources  remain still
either undetermined or not confirmed due to absence or extremal weakness of the spectral lines in these sources.\\
\textbf{Luminosities}: common logarithms of 1.4 GHz isotropic
luminosities are distributed from 39.57 to 42.15 (erg/s) peaking at
$\nu L_r=40.85$ erg/s (Fig1b). As for optical V band luminosities,
they are distributed from $\nu L_o=43.31$ erg/s to $\nu L_o=46.07$
erg/s with a peak at $\nu L_o=44.20$ erg/s (Fig1c). Finally, $\nu
L_x=42.78 - 46.21$ erg/s for ROSAT band X-ray luminosity that peaks
at $\nu L_x=45.10$ erg/s (Fig1d). Radio/optical/X-ray luminosities
are found to be correlated with the redshift above the 99\%
confidence level: correlation coefficient $r$ is equal to 0.53,
0.47, 0.58, respectively (Fig2a,b,c). In that case, we deal with an
evolution from distant ellipicals with powerfool nuclei to normal
elliptical galaxies. However, there is a significant scatters in the
correlations that are probably caused by the scatter in jet
directions (leading to the scatter in Dopper boosting of the
emission) and different brightness state. \\
\textbf{Hosts/nuclei}: up to now, the hosts only of 94 XBLS are
detected. As a rule, they are elliptical galaxies with effective
radii $r_{eff}=3.26 - 25.40$ kpc and ellipticity $\epsilon=0.04 -
0.52$.R band absolute magnitude range from -21.11 mag to -24.72
(Fig1e). The distribution peaks at $M_R=-22.80$ mag and its mean is
-22.83 mag. The nuclei show much broader range of R band absolute
magnitudes - they are distributed from -19.21 mag to -27.24 mag with
a peak at $M_R=-22.20$ mag (Fig1f). There is a negative correlation
with a redshift ($r=-0.52$)above 99\% confidence level (Fig2e)
indicating thus a trend of increasing luminosities towards greater redshifts. \\
\textbf{V-R indices of the hosts}: this quantity ranges from 0.61 to
1.52 and show three different peaks at V-R=0.73, V-R=1.08, and
V-R=1.50 (Fig1g). If confirmed, this may related to the three
"waves" in a birth of elliptical galaxies. However, it may be caused
by poor dataset  - V-R indices are available for only 59 sources.
They show a strong positive correlation with the redshift (r=0.97,
>99\% confidence level; Fig2f) explained with a redshifted emission of
passively evolving elliptical galaxies with an old stellar
population (see \cite{U00}). This correlation, fitted well with a
third-order polinom (see \cite{K11}) may be used to evaluate the
redshifts of that XBLs whose V-R indices are derived photometrically
but their z values remain either unknown or controversial.\\
\textbf{Masses of central BHs}: according to the widely accepted
scenario, BLLs  consist of supermassive BHs in their nuclei whose
masses are estimated mainly via the velocity dispersions in their
hosts. $log M_{BH}/M_\odot$ values are currently estimated for 47
XBLs which range from 7.39 to 9.30 (Fig1h). A peak of the
distribution is found to be at $log M_{BH}/M_\odot=8.30$. These
values do not show a correlation with a redshift. However, there are
weak but statistically significant correlations between $log
M_{BH}/M_\odot$ and 1.4 GHZ/V band luminosities (Fig2g,h).\\
\textbf{Synchrotron peak frequencies}: peak frequencies of
synchrotron SEDs (radio to UV/X-ray frequencies) are currently
determined for 187 XBLs. They range from $log\nu_{peak}=14.56$ Hz to
$log\nu_{peak}=21.46$ Hz with a peak of the distribution at
$log\nu_{peak}=16.60$ Hz (Fig1i). It seems that a subclass of
ultra-high peaked BLLs (UHBLs) with $log\nu_{peak}>19.00$ should be
an artifact of the poor datasets of multifrequency flux values used
for constructing the SEDs of these sources. Among 22 UHBLs, provided
in \cite{N07}, 13 sources are proven to have much lower peak
frequencies in \cite{K11}.  ROSAT band X-ray luminosity shows a
positive correlation ($r=0.40$) with synchrotron peak frequencies
while those for 1.4 GHz and V band do not reveal any trend
(Fig2j,k,m, respectively). This means a trend of increasing
bolometric luminosity towards greater $log\nu_{peak}$ values. This
result is in contradiction the picture of \cite{F98} about the
desreasing power along the sequence LBLs $\rightarrow$ IBLs
$\rightarrow$ HBLs that is explained in \cite{C02} as a cosmological
result of gradual depletion of circumnuclear material
causing a decreasing jet power along this sequence.\\
\textbf{Broadband spectral indices}: radio-to-optical indices are
distributed from 0.17 to 0.59, peaking at $\alpha_{ro}=0.40$. The
range of optical-to-X-ray indices is much broader: $\alpha_{ox}=0.56
- 1.48$ with a peak of the distribution at $\alpha_{ox}=0.98$. As
for radio-to-X-ray indies, they span from $\alpha_{rx}=0.41$ to
$\alpha_{rx}=0.75$ and peak at $\alpha_{rx}=0.56$. The corresponding
figures are Fig1j, Fig1k, and Fig1m, respectively.\\
\textbf{Flux variability}: Almost 60\% of the XBLs are not
investigated for multivavelength flux variability. The best studied
are only brightest XBLs 1ES 2155-304 (since 1890)and Mrk 421 (since
1900). Only several sources have a history of the observation
greater than three decades. XBLs show basically erratical
variability $–$ changing duration, amplitude and base flux level
from flare to flare. Periodical changes are reported very rarely,
e.g. flares with 420 d period in 1ES 2321+419 through 1.5-12 keV
band was reported in \cite{R09}; quasi-periodical flares of 3.2 yr
duration in 1ES 1028+511 through optical R band, reported in
\cite{K09} etc. There is a trend of increasing overall range of
optical variability towards shorter wavelengths (\cite{K11}):
$\langle\triangle m_R\rangle=1.22$ mag (18 sources),
$\langle\triangle m_V\rangle=1.52$ mag (66 sources),
$\langle\triangle m_B\rangle=1.65$ mag (28 sources). At the
intra-night timescales, XBLs are less active compared to RBLs in the
optical domain: flickerings with  $\triangle m\sim0.1$ mag/night
were recorded for several times while there are much more occasions
and higher amplitudes in the case of RBLs. The reason is yet unclear
and can not be simply related to the different jet orientation to
the observer of these two BLL subclasses that could
lead to the different boosting in  the observed flux.\\

\section*{Summary and Conclusions}
\indent \indent In this talk, I have presented a catalogue of 312
XBLs, updated to mid-2010.  In the future, we may expect that the
number of XBLs will grow on the expense of XBL candidates (106
sources) which may not be assumed as BLL sources mainly due to the
lack of high signal-to-noise ratio optical spectra. XBLs are the
extragalactic sources with $0.031<z<0.702$ and with the common
logarithms of radio/optical/X-ray luminosities of 39-42, 43-45, and
42-46 erg/s order, respectively. These sources show a trend of
increasing luminosity towards distant objects that has a deep
cosmological implication - there is an evolution from distant
elliptical galaxies with powerful nuclei into the ellipticals
without active nuclei. However, this correlation may be related
simply to the selection effect - great number of distant BLLs may be
exist whose apparent fluxes are below the detection threshold of
current observing technique. XBL hosts are elliptical galaxies with
effective radii of $3.26-25.40$ kpc and $M_R= -21.11- -24.86$ while
the nuclei reveal much broader range of optical absolute magnitude
of (-19.93, -27.24). $V-R$ indices of the hosts reveal a third order
polynomial relationship with $z$. $log M_{BH}/M_\odot$ values range
with almost two order of the masses up to maximum value of  9.30 and
do not show a correlation  with a redshift. But they show positive
correlations with radio and optical luminosities that may serve as
an argument of the Blandford-Znajek mechanism of jet production.
Bolometric luminosities do not show an decreasing trend towards
higher synchrotron peak frequencies, as it was shown in some
previous works, and the blazar sequence may not be simply explained
by the hypothesis od depleting circumnuclear material along this
sequence. As for Synchrotron peak frequencies
$log\nu_{peak}\sim15-21$ Hz with a mean value of 16.76. We are also
far from perfect understanding both of the character of
multiwavelength flux variability and the nature of unstable
processes, responsible for these variations.

\section*{Acknowledgement}
\indent \indent I would like to thank Ganna Ivashchenko for my
invitation. My participation in the conference is supported by the
Shota Rustaveli Science Foundation grant N3/15.


\begin{thebibliography}{3}
{\small
\bibitem{B92} Bade~N., Engels~D., Fink~H. et al. Astron. Astrophys., V.~254, pp.~21-24 (1992)
\bibitem{B00} Beckmann~V. 'Evolutionary behaviour of AGN: Investigations on BL Lac objects and Seyfert II galaxies', PhD Thesis, University of Hamburg Press,
p.~13 (2000)
\bibitem{B78} Blandford~R.\,D., Rees~M.\,J. in Pittsburgh Conference on BL Lac Objects,University of
Pittsburgh Press,pp.~341-347 (1978)
\bibitem{C02} Cavaliere~A., D'Elia~V. Astrophys. J., V.~571, pp.~226-233 (2002)
\bibitem{D04} Della Ceca~R., Maccacaro~T.,Caccianiga~A. et al. Astron. Astrophys., V.~428, pp.~383-399 (2004)
\bibitem{E92} Elvis~M., Plummer~D., Schachter~J., Fabbiano~G. Astrophys. J. Suppl. Ser., V.~80, pp.~257-303 (1992)
\bibitem{F98} Fossati~G., Maraschi~L., Celotti~A. et al. MNRAS, V.~299, pp.~433-448 (19988)
\bibitem{G84} Gioia~I.\, M., Maccacaro~T., Schild~R.\, E. et al. Astrophys. J., V.~283, pp.~495-511 (1984)
\bibitem{G90} Gioia~I.\, M., Maccacaro~T., Schild~R.\, E. et al. Astrophys. J. Suppl. Ser., V.~72, pp.~567-619 (1990)
\bibitem{G91} Giommi~P., Tagliaferri~G., Beuermann~K. et al. Astrophys. J., V.~378, pp.~77-92 (1991)
\bibitem{H29} Hoffmeister~K. Astron. Nachr., V.~236, pp.~233-246 (1929)
\bibitem{K09} Kapanadze~B.\,Z. MNRAS, V.~398, pp.~832-840 (2009)
\bibitem{K11} Kapanadze~B.\,Z. Astron. J., V.~, pp.~ (submitted) (2011)
\bibitem{M82} Maccacaro~T., Gioia~I.\, M.; Zamorani~G. Astrophys. J., V.~253, pp.~504-511 (1982)
\bibitem{M68} MacLeod~J.\,M., Andrew~B.\,H. Astrophys. Lett., V.~1, pp.~243-247 (1968)
\bibitem{N07} Nieppola~E., Tornikoski~M., Valtaoja~E. Astron. Astrophys., V.~445, pp.~441-450 (2007)
\bibitem{O69} Oke~J.\,B., Neugebauer~G., Becklin~E.\,E. Astrophys. J., V.~156, pp.~41-43 (1969)
\bibitem{O74} Oke~J.\,B., Gunn~J.\,E. Astrophys. J., V.~189, pp.~50-53 (1974)
\bibitem{R70} Racine~R. Astrophys. J., V.~159, pp.~99-103 (1970)
\bibitem{R09} Rani~B., Wiita~P.\,J., Gupta~A.\,C.Astrophys. J., V.~696, pp.~2170-2170 (2009)
\bibitem{S68} Schmitt~J. Nature, V.~218, pp.~663-664 (1968)
\bibitem{U00} Urry~C.\,M., Scarpa~R., O'Dowd~M. et al. Astrophys. J., V.~532, pp.~816-829 (2000)
\bibitem{V69} Viswanathan~N. Astropys. J., V.~155, pp.~133-137 (1969)
\bibitem{W94} Wurtz~R.\,E. 'The Host Galaxies and Galaxy Clustering Environments of BL Lacertae
Objects', PhD Thesis, University of Colorado Press (1994)
\bibitem{W84} Wood~K.\, S., Meekins~J.\,F., Yentis~D.\,J. et al. Astrophys. J. Suppl. Ser., V.~56,
pp.~507-649 (1984)
}
\end{thebibliography}
\end{document}